\documentclass[vci-paper]{vciart}
\usepackage{amssymb}
\usepackage{graphicx}
\usepackage{subfigure}

\begin{document}

\begin{frontmatter}

\title{Micromegas TPC studies at high magnetic fields using the charge dispersion signal}

\author[A,E]{M. Dixit},
\author[B]{D. Atti\'e},
\author[A]{A. Bellerive},
\author[A]{K. Boudjemline},
\author[B]{P. Colas},
\author[B]{P. Giganon},
\author[B]{I. Giomataris},
\author[C]{V. Lepeltier},
\author[A]{S. Liu},
\author[D]{J.-P. Martin},
\author[A]{K. Sachs},
\author[A]{Y. Shin},
\author[A]{S. Turnbull}
\address[A]{Carleton University, Ottawa, Ontario, Canada}
\address[B]{DAPNIA, CEA Saclay, 91191 Gif-Sur-Yvette, France}
\address[C]{LAL, Univ Paris-Sud, IN2P3-CNRS, Orsay, France}
\address[D]{Universit\'e de Montr\'eal, Montr\'eal, Quebec, Canada}
\address[E]{TRIUMF, Vancouver, BC, Canada}

\begin{abstract}
The International Linear Collider (ILC) Time Projection Chamber (TPC) transverse space-point resolution goal is 100~$\mu$m for all tracks including stiff $90^\circ$ tracks with the full  $\sim$2~m drift. A Micro Pattern Gas Detector (MPGD) readout TPC
can achieve the target resolution with existing techniques using 1 mm or narrower pads at the expense of increased detector cost and complexity. The new MPGD readout technique of charge dispersion can achieve good resolution without resorting to narrow pads. This has been demonstrated previously
for 2~mm$\times$~6~mm pads with GEMs and Micromegas in cosmic ray tests and in a KEK beam test
 in a 1~T magnet. We have recently tested a Micromegas-TPC using the charge dispersion readout concept  in a high field super-conducting magnet at DESY. The measured
Micromegas gain was found to be constant within 0.5~\% for magnetic fields up to 5~T. With the strong suppression of transverse diffusion at high magnetic fields,  we measure a flat 50~$\mu$m resolution at 5~T over the full 15 cm drift length of our prototype TPC.
\setlength{\unitlength}{1mm}
 \begin{picture}(0,0)
 \put(75,190){\parbox{5cm}{CarletonHP732\\DAPNIA-07-36\\ LAL0721}}
\end{picture}
\end{abstract}

\end{frontmatter}


\section{Introduction}
The Time Projection Chamber (TPC) \cite{Nygren1975} is a leading candidate for the central 
tracker for future International Linear Collider (ILC). The  ILC TPC resolution goal is 
to measure 200 track points with a transverse resolution of about 100~$\mu$m for the full 2~m drift in a 4~T magnet. The resolution goal, near the fundamental limit from statistics and transverse diffusion, cannot be achieved by the traditional  proportional wire/cathode pad TPC design with its intrinsic  ${\mathbf E}\times{\mathbf B}$ and track angle systematic effects.

The recently developed Micro-Pattern Gas Detectors (MPGD), such as the Micromegas \cite{Giomataris1996} and the GEM
\cite{Sauli1997}, have many advantages for the TPC readout.
The MPGD readout requires less mass for construction and it also naturally suppresses the positive ion space charge build up in the drift volume. 
With the narrow transverse diffusion width in a high magnetic field and negligible ${\mathbf E}\times{\mathbf B}$ and track angle effects,  the MPGD-TPC resolution can, in principle, be much better than for the wire/pad-TPC.  
The traditional TPC could, however, still measure the avalanche position along the wire accurately from the computed centroid of signals induced on relatively wide cathode pads.
For comparable position measurement accuracy and only direct charge signals to work with, the MPGD anode pads will need to be much narrower.
This can significantly increase the MPGD-TPC readout channel count which may be difficult to manage for a large detector.

 The new MPGD readout technique of charge dispersion enables one to use wide pads again and still achieve good resolution.
  The principle of charge dispersion and its application to MPGD-TPC readout has been described in References  ~\cite{Dixit2004,Dixit2006}. In a nutshell,  the conventional MPGD anode is replaced by a high surface resistivity thin film laminated to the readout pad plane with an intermediate insulating spacer. The resistive surface forms a distributed 2-dimensional RC network with respect to the readout plane.
The avalanche charge cluster arriving at the anode disperses with the system RC time constant.
With the avalanche charge dispersing and covering a larger area with time, wider pads can be used without loss of accuracy in charge centroid position determination.

 The charge dispersion MPGD-TPC readout concept has been previously demonstrated in cosmic ray tests without a magnetic field and in a beam test \cite{Bellerive2005,Boudjemline2006,url1} at KEK in a 1~T magnet . Both the GEM and the Micromegas TPC readout systems were tested and good resolution achieved with wide pads. 
For zero drift distance, a resolution of 50~$\mu$m was achieved for 2~mm~$\times$~6~mm pads. Pad width was found no longer to be the resolution limiting factor. The dependence of the measured resolution on the drift distance was close to the expectation from transverse  diffusion and electron statistics .

\section{High field cosmic ray TPC tests in the DESY 5 T magnet}
We have recently tested a Micromegas-TPC prototype with charge dispersion readout
at the DESY super-conducting magnet test facility in magnetic fields up to 5~T. 
Cosmic ray data were collected over a period of about four weeks. 

\newpage
A 10~$\times$~10~cm$^2$ Micromegas with a 50~$\mu$m gap was used for the
readout. The readout plane consisted of a matrix of 128 pads in 9 rows.  The central 7 rows with a total of 126 pads, 2~mm~$\times$~6~mm each, were used for tracking. The two outer rows, each with a single  36~mm~$\times$~6~mm pad, were used for triggering. The readout structure comprised of a 1~M$\Omega/\square$  surface resistivity Cermet  (Al-Si alloy) coated 25~$\mu$m Mylar film  laminated to the readout plane with a 50~$\mu$m thick insulating adhesive. The maximum TPC drift length was 15.7~cm.

\begin{figure}[tb]
\centering
\includegraphics[width=\linewidth]{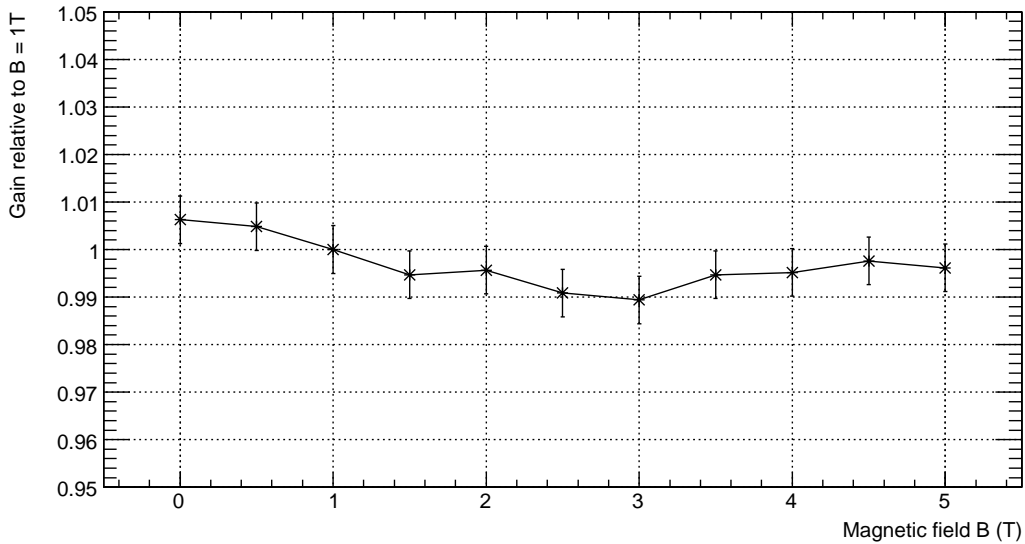}
\caption{Micromegas gain stability from 0 to 5~T for the TPC charge dispersion readout
for Ar:iC$_4$H$_{10}$/95:5 gas mixture.}
\label{figure1}
\end{figure}

The stability of gain as a function of magnetic field was measured first using 5.9~keV $^{55}$Fe x ray source. The Micromegas gain was found to be constant to within 0.5~\% for magnetic fields from 0 to 5~T (Fig.~\ref{figure1}).

For cosmic ray resolution studies, pad signals were read out using ALEPH wire TPC charge preamplifiers and digitized directly without an intermediate shaper amplifier using
200 MHz 8 bit FADCs. Since charge dispersion pulses are slow and signals were integrated over a few
hundred ns during analysis, slower 25 to 40~MHz FADCs would have been adequate.

Two different gas mixtures were tested. The first, Ar:iC$_4$H$_{10}$/95:5, was chosen as reference
to compare with our previous KEK measurements at 1 T.  The second, so called T2K gas,  Ar:CF$_4$:iC$_4$H$_{10}$/95:3:2, is  a possible candidate for the ILC TPC. It has a high 73~$\mu$m/ns electron drift velocity  at a moderate 200~V/cm electric field, a relatively low longitudinal diffusion of D$_{\rm L} \simeq 248~\mu{\rm m}/\sqrt{\rm cm}$, and  a large $\omega\tau\sim~20$ at 5~T, which reduces transverse diffusion to D$_{\rm Tr} \simeq 19~\mu{\rm m}/\sqrt{\rm cm}$.

The transverse TPC resolution was measured at 5 T for both gases to benchmark TPC performance in a magnetic field of strength comparable  to that for the ILC detector. For Ar:CF$_4$iC$_4$H$_{10}$/95:3:2, resolution measurements were also carried out at 0.5 T  to measure the effect of diffusion.

\section{Cosmic ray resolution measurements in magnetic field}
The data analysis was carried out following procedure described previously \cite{Bellerive2005,Boudjemline2006}.  
A subset of data was used for calibration and the remaining data used to measure the resolution.
 The calibration data set was used to determine the pad response function (PRF) and also  to determine the bias correction. The bias is a systematic error in the absolute position measurement. It is attributed to point-to-point variations in the capacitance per unit area and the surface resistivity for the anode RC structure. Since the bias is intrinsic to the detector and does not change with time,  a bias correction can be reliably applied. An initial bias of up to 100~$\mu$m was observed.  After the correction, the bias remaining was small, about $\pm 20~\mu$m.
 
   A momentum cut has been applied to eliminate tracks with large angles with respect to pads.  The cut removed tracks with  $p_{T} < 2$~GeV/c at 5~T and $p_{T} < 0.3$~GeV/c at 0.5~T.
 As in our previous  work, the resolution measurements reported below are for track angles   $|\phi|<5^\circ$.

\subsection{Resolution at 0.5~T as a function of gas gain}
\begin{figure}[tp]
\centering
\subfigure[]{\includegraphics[width=0.49\textwidth]{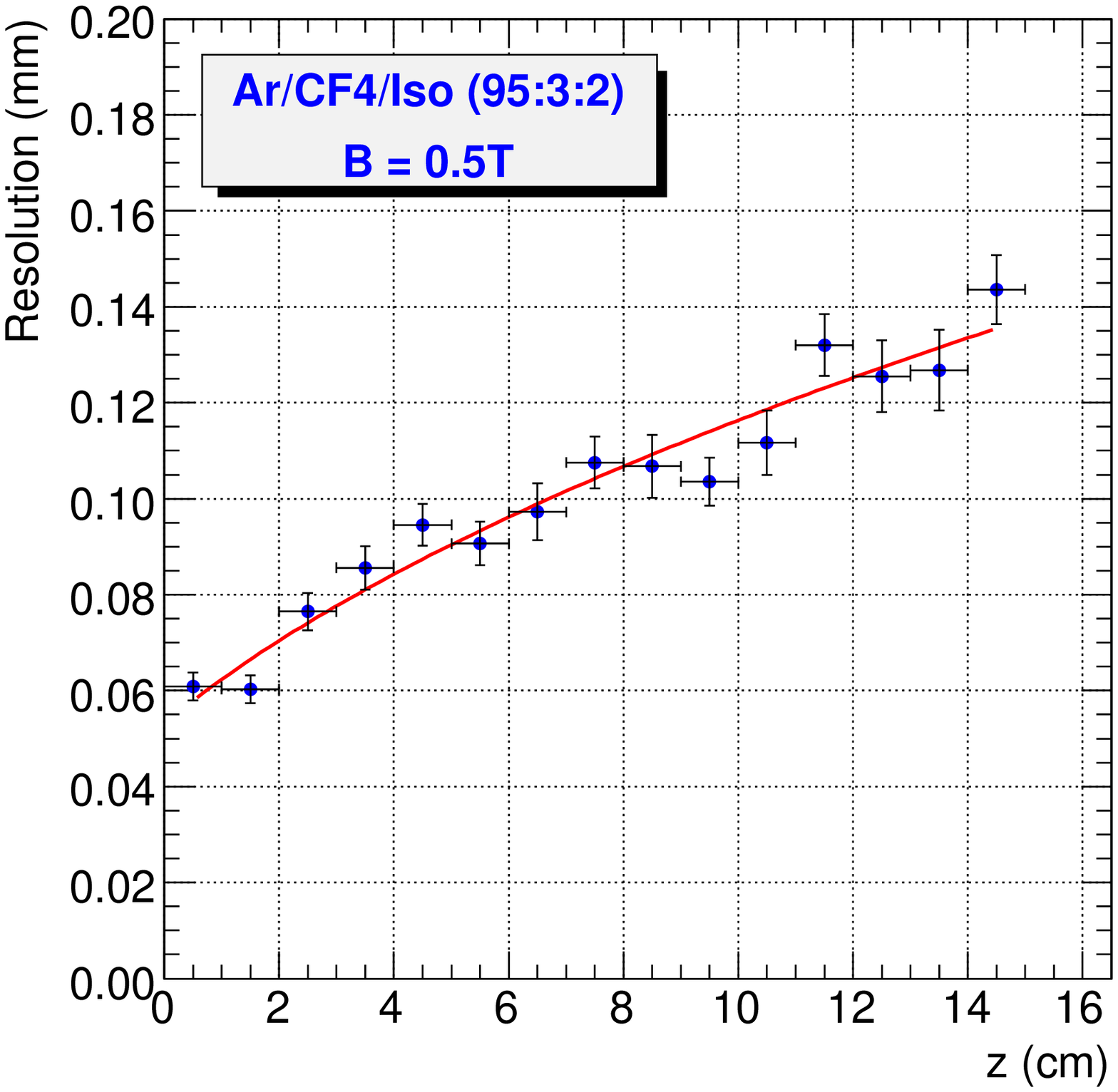}} 
\subfigure[]{\includegraphics[width=0.49\textwidth]{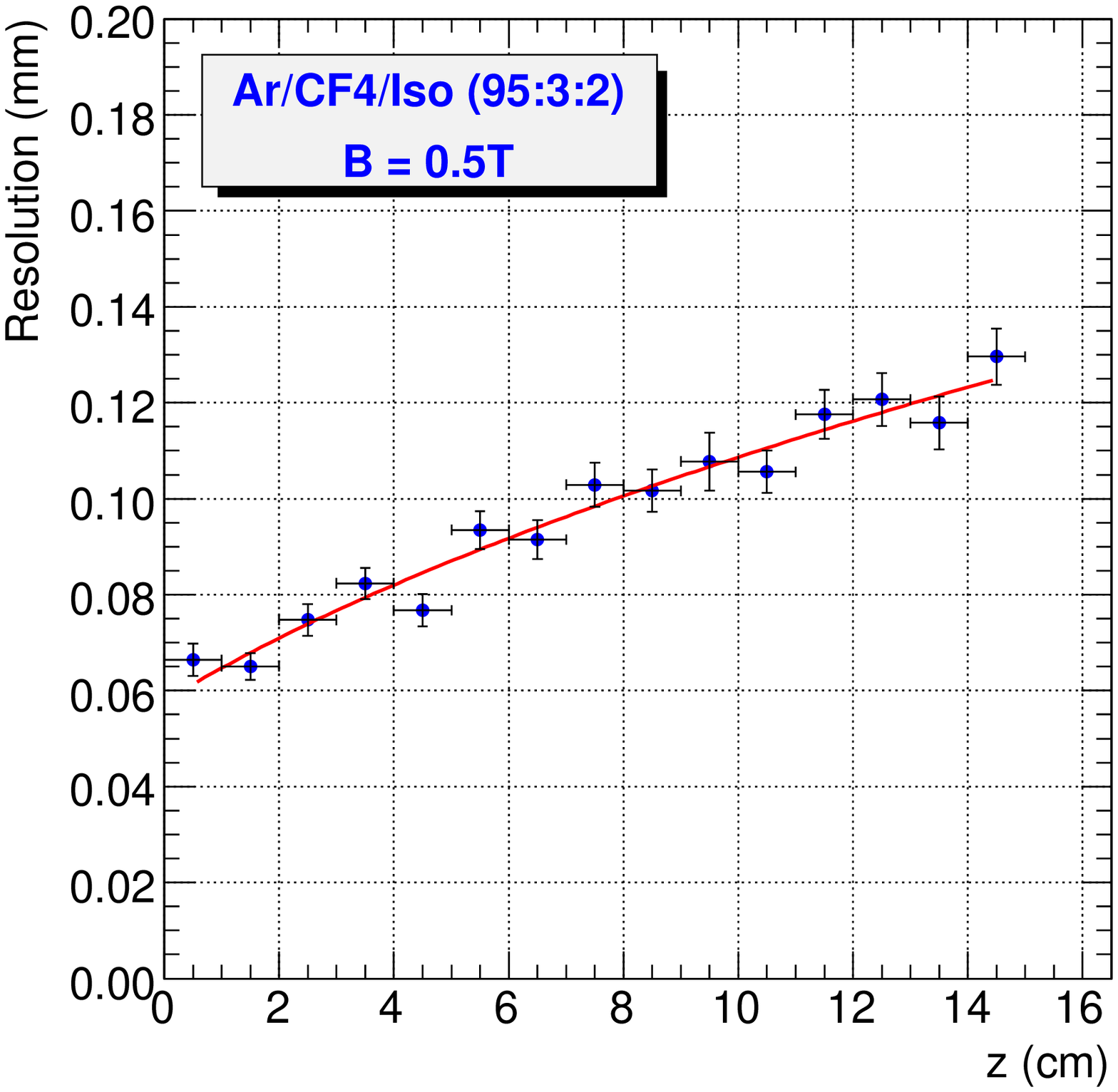}}
\caption{Transverse resolution as a function of drift distance for  2~mm x 6~mm
         pads using the Ar:CF$_4$:iC$_4$H$_{10}$/95:3:2 gas mixture for magnetic
         field of 0.5~T at different gains; (a) 4700 and (b) 2300. 
         The solid line is fitted to the resolution expected from diffusion in 
         the TPC gas and electron statistics (Eq.~\ref{eq:reso})}
\label{figure2}
\end{figure}

The resolution is given by the geometric mean of standard deviations of residuals from track fits
done in two different ways: including and excluding the row of pads for which the resolution is being
determined. The measured resolution  is shown in Fig.~\ref{figure2}
 for Ar:iC$_4$H$_{10}$/95:5 gas mixture as a function of the drift distance $z$ for two different gains.
The $z$ dependence of resolution is given by:

\begin{equation}
\sigma = \sqrt{\sigma_0^2 + \frac{D_{\rm Tr}^2{z}}{N_{\rm eff}}} \; ,\label{eq:reso}
\end{equation}

where $\sigma_0$ is the resolution at zero drift distance and $D_{Tr}$ is the transverse diffusion constant. Here 
$N_{\rm eff}$ is the effective number of  electrons over the length of a pad.  
$N_{\rm eff}$ is not the average number of electrons, but is given by:
$N_{\rm eff} ={1/ \langle \sqrt{1/N}  \rangle}^2$,
where N is the number of electrons following the Landau distribution and including the effects of gas gain fluctuations. $N_{\rm eff}$ was obtained from the resolution fit using the transverse diffusion constant $D_{Tr}$ as calculated by Magboltz \cite{Magboltz}.

At a gas gain of $ \simeq 4700$, the measured resolution at zero drift distance $\sigma_0$ was  about $50~\mu$m  with N$_{\rm eff}=25.2\pm 2.1$. A lower gain TPC operation without sacrificing resolution  would be desirable since it minimizes space charge effects from the buildup of positive ions  in the drift volume. Resolution  measurement at a gain of $\simeq 2300$ at 0.5~T found the same $\sigma_0$, but with N$_{\rm eff}=28.8\pm 2.2$.  The mean value for N$_{\rm eff}$ for the two measurements was  $26.8\pm 1.5$. The  TPC resolution was obviously  not compromised by the lower gain operation.

\subsection{Resolution at 5~T as a function of the gas mixture}
\begin{figure}[htp]
\centering
\subfigure[]{\includegraphics[width=0.49\textwidth]{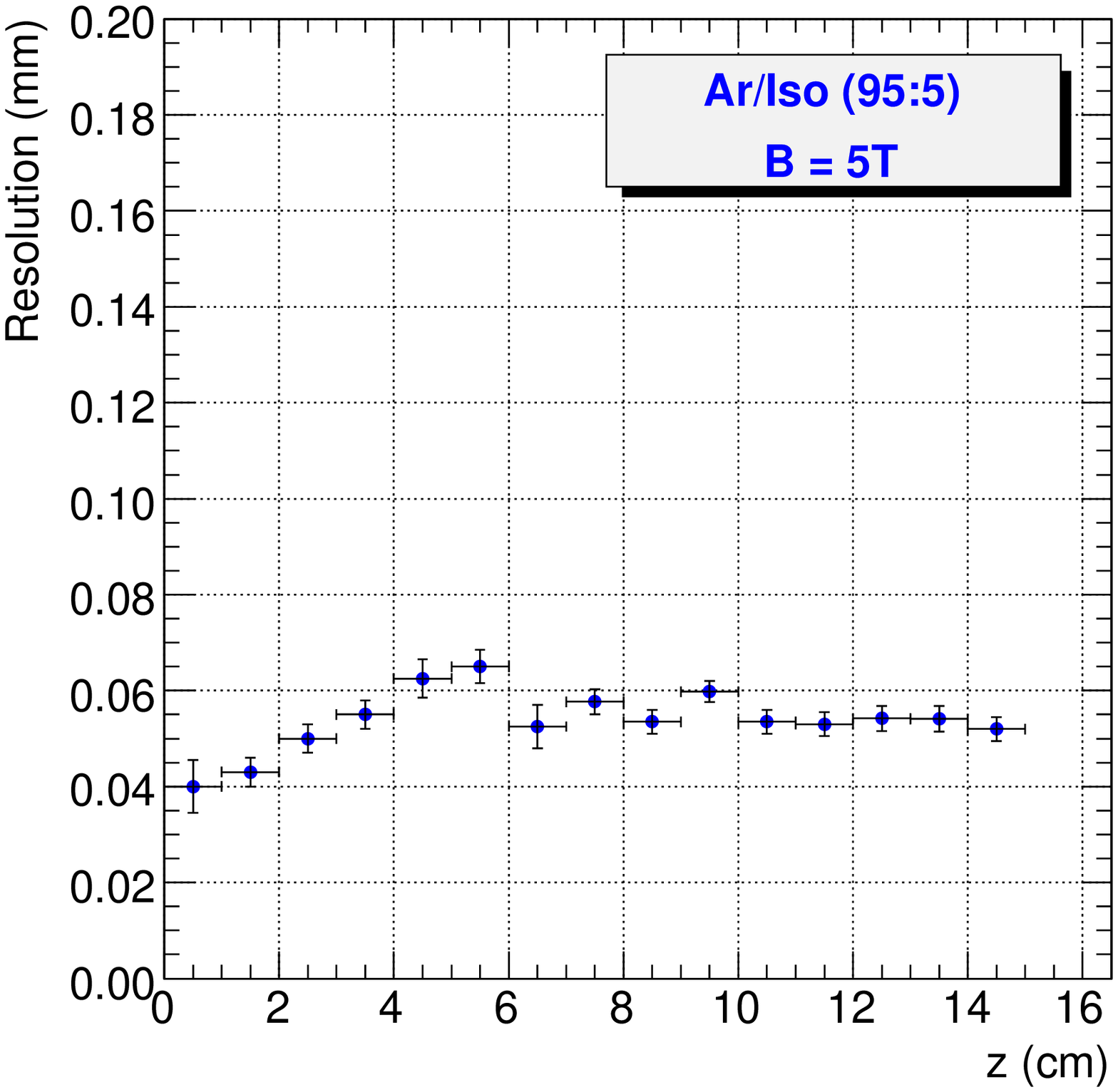}} 
\subfigure[]{\includegraphics[width=0.49\textwidth]{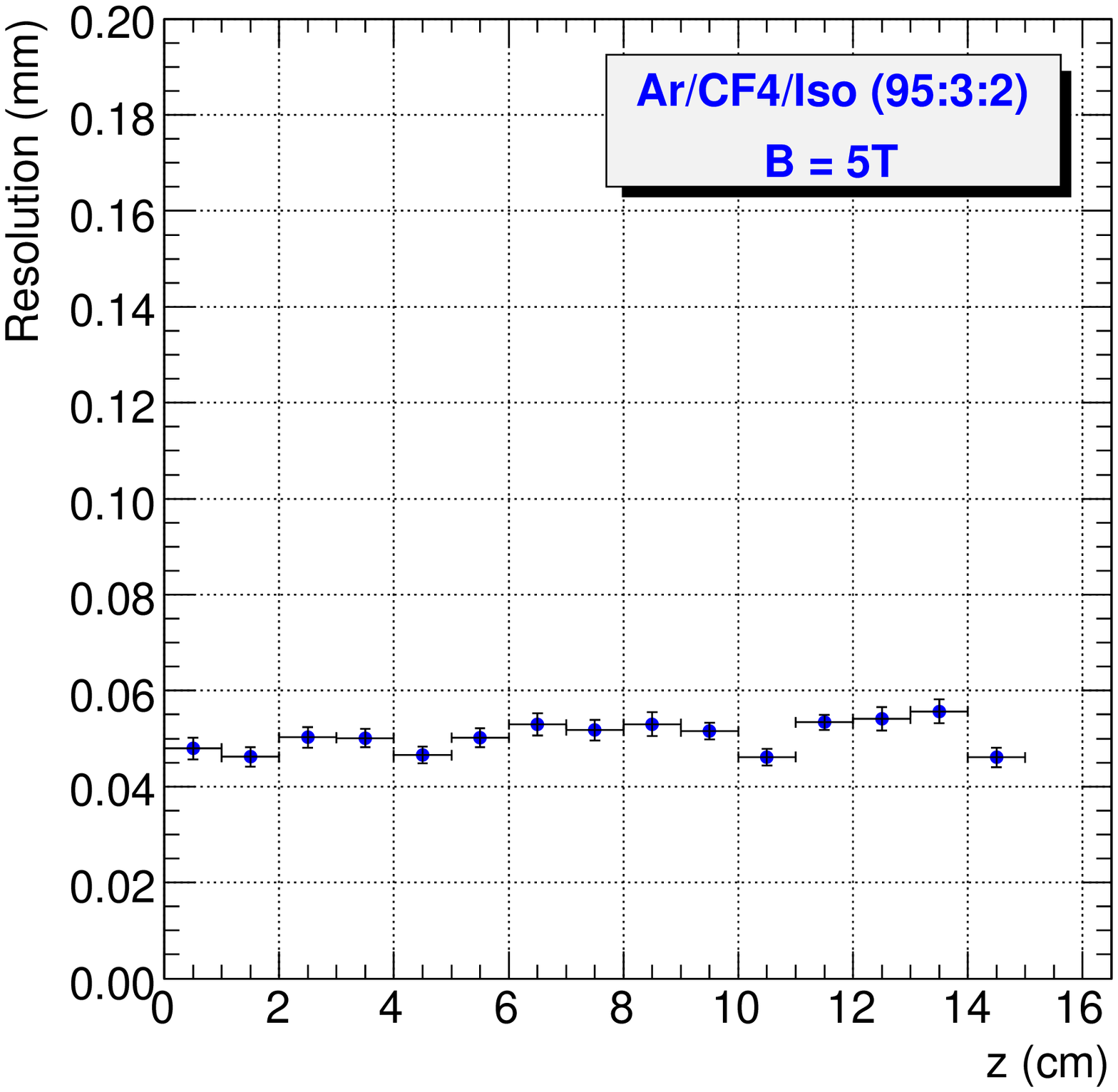}}
\caption{Transverse resolution as a function of drift distance for  2~mm x 6~mm
         pads for a magnetic field of 5~T for two gases mixtures: 
         (a) Ar:iC$_4$H$_{10}$/95:5  and (b) Ar:CF$_4$:iC$_4$H$_{10}$/95:3:2.}
\label{figure3}
\end{figure}

Fig.\ref{figure3} shows resolution as a function of the drift distance for the two different gas mixtures for which
the effect of diffusion should be negligible over the 15.7 cm TPC drift length. The measured resolution was found to be independent of $z$ for these measurements; i.e. $\sigma(z)_{[B=5{\rm T}]} \simeq \sigma_0$. 

\section{Summary and outlook}
With a charge dispersion readout, the Micromegas-TPC gas gain was found to be stable in magnetic fields up to 5~T. The transverse resolution was also measured. With charge dispersion, the  resolution is no longer limited by the width of the readout pads. A flat  $\sim 50~\mu$m resolution was achieved with 2~mm x 6~mm pads at 5~T over the full 15 cm TPC drift length. At a magnetic field of 0.5~T, the resolution was measured at two different gas gains. The  TPC resolution was  not compromised by the lower gain operation. The dependence of resolution on drift distance was as expected from diffusion. 
These measurements are significantly better than has been previously achieved with conventional direct charge MPGD-TPC readout techniques.
The extrapolation to the ILC TPC conditions is promising. With good control of systematics,  a TPC resolution better than $100~\mu$m should be achievable over the entire $\sim$2~m long drift region with 2~mm x 6~mm pads at a gain of $\sim 2000$ in a 4-5~T field.

\section{Acknowledgments}

We wish to thank Ron Settles for lending us the ALEPH TPC wire charge preamplifiers and Rolf Heuer and Ties Behnke for facilitating access to the 5~T super-conducting magnet test facility at DESY for these measurements. 
We thank Vance Strickland and Matt Bowcock at Carleton
for helping sort out mechanical details and Philippe Gravelle for his careful work in solving a variety of technical problems. The research was supported by a project
grant from the Natural Science and Engineering Research Council of Canada and the Ontario Premier's Research Excellence Award (IPREA). Partial support by DESY is also gratefully acknowledged. TRIUMF receives federal funding via a contribution agreement 
through the National Research Council of Canada.



\newpage
\begin{thebibliography}{99}
\bibitem{Nygren1975} D.R. Nygren, A time projection chamber—1975, Presented at 1975 PEP Summer Study, PEP 198, 1975 and included in Proceedings.
\bibitem{Giomataris1996} Y. Giomataris, et al., Nucl. Instr. and Meth. A 376 (1996) 29.
\bibitem{Sauli1997} F. Sauli, Nucl. Instr. and Meth. A 386 (1997) 531.
\bibitem{Dixit2004} M. S. Dixit et al., Nucl. Instrum. Meth., A518 (2004) 721.
\bibitem{Dixit2006} M. Dixit and A. Rankin, Nucl. Instrum. Meth. A566 (2006), 28.
\bibitem{Bellerive2005}A. Bellerive et al., Spatial Resolution of a Micromegas TPC Using the Charge Dispersion Signal, Proceedings of International Linear Collider Workshop, LCWS2005, Stanford, USA,arXiv:physics/0510085.
\bibitem{Boudjemline2006} K. Boudjemline et al., Spatial resolution of a GEM readout TPC using the charge dispersion signal, accepted for publication, Nucl. Instrum. Meth. A, arXiv:physics/0610232.
\bibitem{url1} http://www.physics.carleton.ca/research/ilc/presentations.html.
\bibitem{Magboltz} S. Biagi, Magboltz 2, version 7.1 (2004) CERN library.


\end{thebibliography}
\end{document}